# A simple method of measuring profiles of thin liquid films for microfluidics experiments by means of interference reflection microscopy

V. Berejnov and D. Li

*Department of Mechanical and Mechatronics Engineering, University of Waterloo, 200 University Ave. West Waterloo, Ontario N2L 3G1, Canada*
*Corresponding author: E-mail: berejnov@gmail.com

**Abstract**

A simple method was developed to observe the interference patterns of the light reflected by the interfaces of thin liquid films. Employing a fluorescent microscope with epi-illumination, we collected the 2D patterns of interference fringes containing information of the liquid film topography at microscale. To demonstrate the utility of the proposed visualization method we developed a framework for reconstructing the profiles of liquid films by analysing the reflected interferograms numerically. Both the visualization and reconstruction methods should be useful for variety of microfluidic applications involving the flows with droplets and bubbles in which the knowledge of the topography of the interfacial liquid film is critical.

**Introduction**

Visualization and interpretation of the topography of liquid films is an essential and challenging part of research for numerous fundamental problems and applications [1-3] regarding the interfacial phenomena. The importance of this subject is supported by unique sensitivity of a liquid interface to mechanical perturbations and chemical doping. The liquid film easily adjusts its shape in response to small disturbances. This property turns the thin film into a sensitive indicator able to visualize the interplay between delicate physical and chemical effects and processes happening on the small scale of the order from ~10 nm to ~100 μm. In many cases[4-12], considering the resultant profile of the thin liquid film in response to particular conditions of disturbance, it is possible to conduct a backward analysis and obtain a correlation: film shape versus strength of mechanical, physical, or chemical perturbation on the scale of the film thickness. While the correlation is established a thin film could be used as an indicator of microscopic forces, shears, and pressures.

The non-invasive optical methods are very suitable for characterizing geometry of the liquid film since they have the least disturbance to the interface. The methods based on interferomery are the most successive among others optical techniques since they employ the wavelength of light as natural length scale for reconstruction of the film topography making no needs in additional calibration. To date several techniques were developed for examining one[13-15], two[7, 11, 12, 16], and three[4, 5, 8] dimensional profiles of the thin liquid films for different applications. One may expect that by employing similar interferometry methods for acquiring and reconstructing the thin film topography, a variety of fundamental and application studies at microscale could be potentially benefitted: hydrodynamics of motion of drops and bubbles in tubes and their coalescence[17-20], mechanics of foams and liquid membranes[9, 21], physical chemistry of capillarity and liquid/solid interaction including disjoining effects[2, 22, 23], electrokinetics of multiphase flows in microchannels[24-26], and adhesion of biological cells to the liquid/solid interface[4].

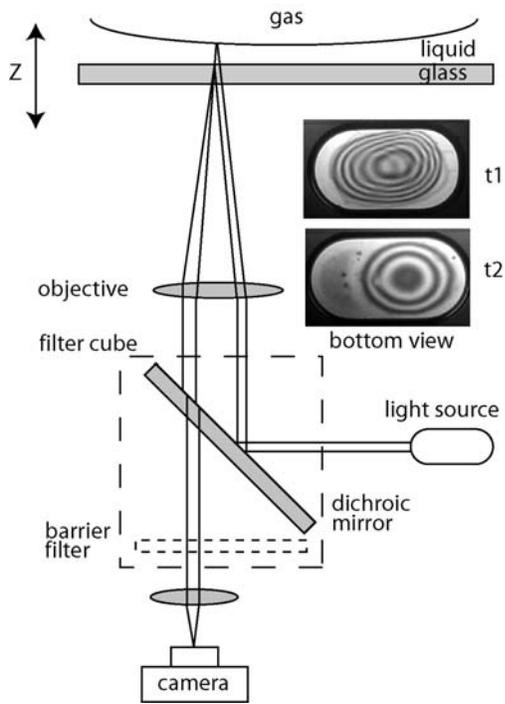

**Figure 1** Optical schematic of the modified microscope. The filter cube is shown with the removed barrier filter – dashed rectangle. Light source is a halogen or Hg lamp. Interference pattern is located near the glass/liquid interface. Objective is focused on the interference pattern due to adjustment of the Z axis. The insert is the bottom view of the interference fringes for two different times t1<t2. The bubble is in the rectangular microchannel; focus is on the liquid/glass interface; the patterns with interference fringes are localized in the same focal plane.

However, despite the detailed shape of the thin film that optical interference methods could deliver in principle, these methods are still lack of the application universality and user-simplicity. They are not friendly in operation, often require both specific tools and experience for data acquisition and interpretation, and not suitable for liquid films. Up to date, a standard method of interference reflection microscopy, IRM, is applied for examining the topography of the solid films. However, the *thin liquid films* which are critical for microfluidics applications lack of the standard optical tool for analysing their profiles.

In this Letter, we introduce a simple method of converting the common fluorescent microscope with epi-illumination into the IRM mode. The patterns of interference fringes of equal thickness of liquid film could be collected routinely at standard conditions. To develop a fully integrated analytical tool we also created a method for analysing the interference fringe patterns that outputs the feasible profiles of the measured liquid films.

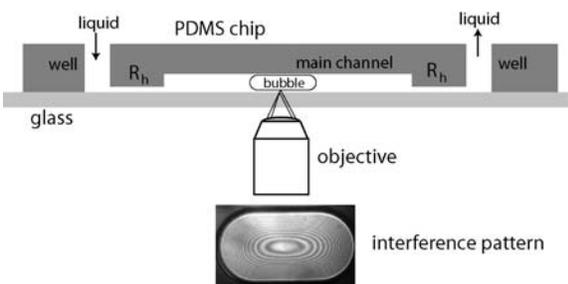

**Figure 2** Two layered microfluidic chips, side view. $R_h$ denotes the channels supporting the hydraulic resistance. For chip geometry see the text. In the monochromatic light the bubble film, previously disturbed, produces the interference pattern with

**Experimental**

In our experiments we used a standard fluorescent microscope Nikon Eclipse Ti with epi-illumination and automatic Z-drive. All tests were performed with the objective 10x0.3 Plan Fluor, Nikon Inc, distance from the objective to the sample was 2.9 cm, the estimated depth of focus, DOF, was 7.2 μm for $\lambda$=650 nm. We tested two kinds of illuminations: from the halogen lamp and from the high pressure Hg lamp installed in the Nikon Intensilight C-HGFIE illumination device. The microscope was used in standard fluorescent mode with modified fluorescent cube CY5 HYQ. As illustrated in Fig. 1, the emission barrier filter in the cube was removed, while the dichromatic mirror (that is a beam-splitter) and the excitation filter were kept in the cube. This conversion allows delivering to the sample and from the sample the monochromatic light with appropriately narrow spectral range. The cube CY5 HYQ has a wide 60-nanometer excitation band covering a spectral range from ~575 to 650 nm matching the 577.-579.0 nm peaks of the Hg lamp. We have found little difference in contrast between halogen-lamp and Hg-lamp illuminations.

Interference patterns from the sample were collected with the black/white CCD camera Retiga 200R, QImaging Inc, capable to record 8 bit and 12 bit images at 2x2, 4x4, and 8x8 binning. The tested interval of exposure from 1 msec to 1sec delivered contrast images of the interference patterns. Images were pre-processed with the software NIS-Elements BR 3.00 SP4, Imaging Software Inc. The presented data corresponds to the raw images having dimension of 800x600 pixels with 8 bit resolution of grey color. The calibration micrometer/pixel was 2.11 μm. The raw images were recorded with 1 msec of exposure time with 2x2 binning.

The sample was a rectangular microchannel fabricated in the PDMS/glass microfluidic chip, as shown in Fig.2. The channel was filled with the Borate buffer (pH=9.5, 7.5mM); controlling the buffer injection we were able to manage the size of the air bubble. We found that both the composition of liquid and the PDMS/glass treatment strictly affect the thickness and the properties of liquid film. The microfluidic chip, Fig. 2, was fabricated employing standard two layers soft-lithography technique[27, 28]. The length, depth, and width of the main rectangular channel, Fig. 2, were 25 mm, 70 μm, and 200 μm, respectively. The main channel was terminated with two additional channels: the length, depth, and width of which were 3 mm, 5 μm, and 500 μm, respectively. The purpose of these small channels was to provide hydraulic resistance against the pressure induced flows and isolate the bubble from the wells due to capillary forces.

A PDMS polymer, Sylgard 184, Dow Corning, was cast against the two layers master. After the PDMS was cured (during 24 h at T=80°C), it was cut and peeled away from the master yielding a replica containing a positive structure of the microfluidic channel, then the inlet and outlet ports were punched with rounded holes. The thickness of all replicas was approximately 2~3 mm. The microfluidic chips were assembled from the PDMS replicas and the glass slides, GoldLine Microscope Slides, VWR International. The replicas and glass substrates were RF treated for 1 min using PDC-32G, Harrick Scientific, and irreversibly sealed due to immediate conformal contact. Then, the assembled chips were filled with the Borate buffer (pH=9.5, 7.5mM) and used for tests after 2 hours. The air bubble was introduced in the filled microfluidic channel to provide a thin liquid film.

The loaded sample was installed on the microscope stage and the bottom part of the bubble interface, a liquid film on a glass substrate, was positioned in focus (the contrast of the reflected interference patterns for air/liquid/glass interfaces was better than for air/liquid/PDMS interfaces). Then, the small pressure difference was applied to let the bubble move slightly in one direction. Immediately the interference fringes, Fig. 1, appeared visualising the disturbance of the initially flat liquid film of the Borate buffer on a glass substrate. We begun observation at time t1, (Fig. 1), when the shape of the film resembled a well formed dimple[29, 30]. We observed that the bubble tends to squeeze the liquid from the film to reduce the depth of a dimple, (insert t2 on Fig. 1). Eventually, the film restores its initial flat shape. No interference fringes were observed for this final state. The images of the dimple-like fringes were recorded and analysed in order to reconstruct the detailed profile of the film.

**Method of phase reconstruction from interferograms**

The local intensity of the interference pattern is a function of the phase change which, in turn, depends on the local film thickness. Below, we present a theoretical framework of the film profile reconstruction which avoids the detailed analysis of the intensity pattern and fringe indexing with respect to the order of the phase change. For our experimental conditions[31] we can ignore an effect of the objective aperture on the intensity of interference. Intensity of light, *I*, reflected from the thin liquid film interfaces depends on the film thickness, *h*, which is a function of the (*x*, *y*) coordinates. The general expression[10, 32] for *I(x, y)* at any point (*x, y*) can be written in the following form:

$$I = \frac{\alpha + \beta \cos(2\theta_l)}{\kappa + \beta \cos(2\theta_l)},  \qquad \text{Eq. 1}$$

where the $\theta_l$ is a phase change. The phase change depends on the film thickness

$$\theta_l = \frac{2\pi n_l h}{\lambda},$$  Eq. 2

where $\lambda$ is a wavelength of light, and $n_l$ is a refractive index of liquid, Fig. 1. Parameters $\alpha$, $\beta$, and $\kappa$ are the material coefficients depending on the refractive indexes of liquid, $n_l=1.33$, vapour, $n_v=1.0$, and solid, $n_s=1.55$, respectively, and could be computed by employing the following expressions:

$$\alpha = r_1^2 + r_2^2, \quad b = 2r_1 r_2, \quad \kappa = 1 + r_1^2 r_2^2,$$  Eq.3

were $r_1$ and $r_2$ are the Fresnel coefficients:

$$r_1 = \frac{n_l - n_v}{n_l + n_v}, \quad r_2 = \frac{n_s - n_l}{n_s + n_l}.$$  Eq. 4

The recorded, 8 bit, image is an array of the pixels in which intensity values represented by grey color in an interval from 0 (black) to 255 (white). The interferograms located in the region of interest, ROI, is the 2D grey array G(x, y), where x and y are the pixel indexes in the ROI. The normalized intensity, $\bar{I}(x,y)$ of the reflected light is proportional to the normalized grey function $\bar{G}(x,y)$. For simplicity we limit our analysis to the cases $\bar{G}(x, y = const) \equiv \bar{G}(x)$ and $\bar{I}(x, y = const) \equiv \bar{I}(x)$, where the condition $y = const$ corresponds to a raw of pixels taken from the array G(x, y). The expression for normalized intensity[10, 32] is

$$\bar{I}(x,y) = \frac{I(x,y) - I_{min}(x,y)}{I_{max}(x,y) - I_{min}(x,y)},$$  Eq. 5

where maximal and minimal values correspond to the white and black fringes on the interference pattern and are defined as $I_{max} = \frac{\alpha + \beta}{\kappa + \beta}, I_{min} = \frac{\alpha - \beta}{\kappa - \beta}$, respectively.

Substituting all of the above expressions in the equation, Eq. 1, for $I(x)$ and solving it with respect to $\cos(2\theta_l)$ yields the following formula for the film profile:

$$\cos(2\theta_l) = \frac{\beta + k - 2\kappa \bar{G}}{-\beta - k + 2\beta \bar{G}}[33].$$  Eq. 6

The right part of Eq. 6 is based entirely on the experimental data, while the left part is a function of the film thickness h(x). For computational purpose it is not worth to invert the above expression solving it with respect to $2\theta_l$. This operation delivers the periodic form of $2\theta_l$ while the monotonous form for h(x) is expected. In order to convert this periodic form into monotonous the image of fringes needs to be processed through a digital analysis indexing every fringe with respect to the order of interference. Then, Eq. 6 can be applied for the given maximum-minimum interval (white-black fringe pattern) and, thus, the phase change for the fringe pattern interval can be calculated. Next, collecting the phase changes for indexed fringes we could construct the whole function of the phase with respect to the coordinate $\theta_l(x)$. While the $\theta_l(x)$ is obtained the thickness h(x) can be found from Eq. 2. Below we present another method for direct evaluation $\theta_l(x)$, which does not require the digital image analysis and employs an integral representation of the phase change.

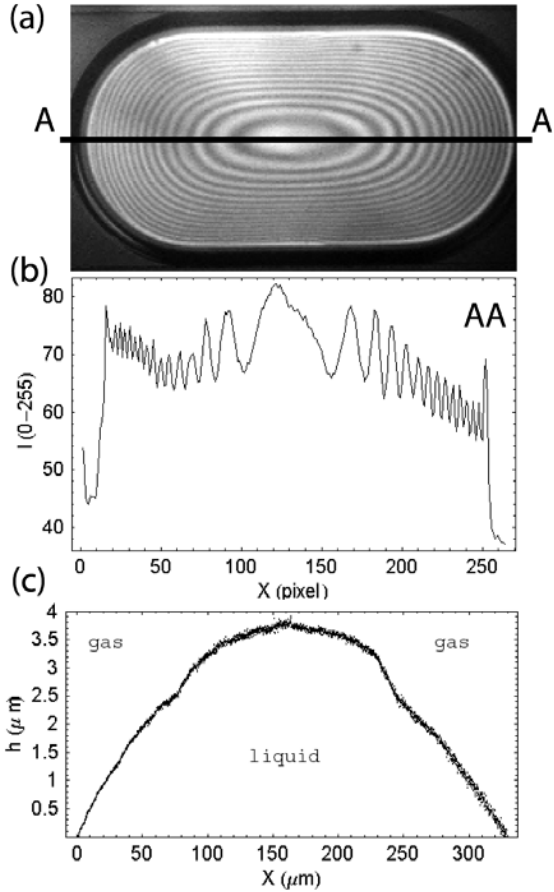

**Figure 3** Reconstruction of the liquid film profile from the interference pattern. (a) is the raw image of the interference pattern, (b) is a raw grey function G(x, y=125), and (c) is a reconstructed dimple-like profile of the liquid/gas interface.

**Direct method of phase reconstruction**

Let's denote $\cos(2\theta_l(x)) \rightarrow F(x)$, where F(x) is the generalized and interpolated experimental data. The derivation of this expression with respect to x gives:

$$-2\theta_l^{'}(x)\sin(2\theta_l(x)) = F^{'}(x).$$  Eq. 7

Employing an identity: $\sin^2(y) + \cos^2(y) = 1$ for these two expressions yields a differential equation for $\theta_l(x)$:

$$F^2(x) + \left(\frac{F^{'}(x)}{-2\theta_l^{'}(x)}\right)^2 = 1.$$  Eq. 8

Solving this equation with respect to the phase term $\theta_l(x)$ and then performing an integration gives us an expression for $\theta_l(x)$ which is not transcendental:

$$\theta_l(x) = \int \frac{F'(x)}{2\sqrt{1-F(x)^2}}dx. \qquad \text{Eq. 9}$$

The functions $F'(x)$ and $F(x)$ are known from the experimental image, therefore the phase $\theta_l(x)$ can be computed directly from the last expression, Eq. 9. Then, te profile of the liquid film, $h(x)$, can be calculated from Eq. 2.

**Results**

The following procedures were performed for obtaining the profile of the dimple presented on Fig.3. We acquired the raw data $G(x, y)$; subtracted one dimensional array y=C from $G(x, y=C)$; averaged this array with its neighbours $<G>|_{y-2,y-1,y,y+1,y+2}$ to reduce the data noise; interpolated the averaged data extending the number of points to increase accuracy of further integration. Then we applied convolution for $<G>$ and obtained a median function of the $<G>$ data; we computed normalized function $\overline{G}(x) = \frac{<G>(x) - <G>_{min}(x)}{<G>_{max}(x) - <G>_{min}(x)}$ and calculated the integral for the phase $\theta_l(x)$ and then the film thickness $h(x)$. The result of the above framework is presented on Fig.3, where we reconstructed the dimple profile (panel c) employing the pattern of the interference fringes (panels a and b) recorded by our optical method.

**Summary**

In summary, we report *i*) a method of interferometric visualization of the topography of the thin liquid films using a standard fluorescent microscope equipped with epi-fluorescent illumination and *ii*) a theoretical framework suitable for reconstructing the liquid film profile from the recorded interference pattern. Because only the standard components are required this method would be very useful for analysis of the conditions of motion of the bubbles and drops in the microfluidic channels at different conditions.

The authors wish to thank the financial support of the Canada Research Chairs program (Li), the Natural Sciences and Engineering Research Council through a research grant to D. Li. V.Berejnov acknowledge B. Rubinstein for helpful discussion.